\documentclass[11pt,twoside]{article}
\usepackage{asp2010}
\usepackage{epsf}
\usepackage{lscape}
\usepackage{natbib}
\usepackage{graphicx}

\begin{document}
\bibliographystyle{asp2010}

\title{AGN feedback using AMR cosmological simulations}

\author{Yohan Dubois$^1$, Julien Devriendt$^{1,2}$, Adrianne Slyz$^1$, and Romain Teyssier$^{3,4}$
  \affil{  $^1$Astrophysics, University of Oxford, Denys Wilkinson Building, Keble Road, Oxford, OX13RH, United Kingdom\\
  $^2$CRAL, Universit\'e de Lyon I, 9 Avenue Charles Andr\'e, 69561, St-Genis-Laval Cedex, France\\
  $^3$Universit\"at Z\"urich, Institute f\"ur Theoretische Physik, Winterthurerstrasse 190, CH-8057 Z\"urich, Switzerland\\
  $^4$CEA Saclay, DSM/IRFU/SAP, B\^atiment 709, F-91191 Gif-sur-Yvette, Cedex, FranceÊ}
    }

\begin{abstract}
Feedback processes are thought to solve some of the long-standing issues of the numerical modelling of galaxy formation: over-cooling, low angular momentum, massive blue galaxies, extra-galactic enrichment, etc. The accretion of gas onto super-massive black holes in the centre of massive galaxies can release tremendous amounts of energy to the surrounding medium. We show, with cosmological Adaptive Mesh Refinement simulations, how the growth of black holes is regulated by the feedback from Active Galactic Nuclei using a new dual jet/heating mechanism. We discuss how this large amount of feedback is able to modify the cold baryon content of galaxies, and perturb the properties of the hot plasma in their vicinity.
\end{abstract}

\section{Introduction} 

It is now well accepted that galaxies host super-massive Black Holes (BHs) in their centres that can potentially power large-scale outflows. Observations show tight relationships between the BH mass and their host galaxy properties \citep{magorrianetal98}. \cite{silk&rees98} have proposed that the energy liberated by the accretion mechanism onto BHs is sufficient enough to unbind galactic gas and trigger strong outflows.
This AGN feedback mechanism is confirmed by a large body of evidence through the detection of X-ray cavities and radio jets in galaxy clusters, and Broad Absorption Lines in the spectra of QSOs.

Numerical simulations of galaxy formation suffer from the overcooling problem: too much gas can collapse during a Hubble time and lead galaxies that are too massive and produce too many stars. Semi-analytical models of galaxy formation on top of Dark Matter (DM) only simulations have been successful at fitting the bright-end part of the luminosity function of galaxies if they include an efficient AGN feedback process \citep{boweretal06}.

In \cite{duboisetal11agnmodel} we proposed a dual jet/heating mechanism for AGN feedback together with its BH growth. This is the first time for hydro cosmological simulations using the Adaptive Mesh Refinement (AMR) technique with the RAMSES code \citep{teyssier02}. We review in this paper the main results obtained from these simulations.

\section{AGN feedback model} 

We recall in short the main features and parameters of our model for AGN feedback which are given in more detail in \cite{duboisetal11agnmodel}.

We assume an initial seed mass of $10^5 \rm \, M_{\odot}$ for each BH formed in the centre of a galaxy with sufficient gas density $\rho>\rho_0$ and stars with a stellar fraction $f_s=\rho_s/(\rho_s+\rho)>0.25$. We ensure that only one BH per galaxy is created by forcing BHs to form at a distance $100$ kpc larger that any BH already in the simulation box. BHs are allowed to merge together when they become close enough (distance $r<4 \Delta x$, where $\Delta x$ is the minimum cell size).

BHs can not only grow by successive mergers but also by accreting gas at the following Bondi accretion rate
\begin{equation}
\dot M_{\rm BH}={4\pi \alpha G^2 M_{\rm BH}^2 \rho \over (c_s^2+ u^2) ^{3/2}}\, ,
\label{dMBH}
\end{equation}
and with a maximum accretion rate defined by the Eddington limit
\begin{equation}
\dot M_{\rm Edd}={4\pi G M_{\rm BH}m_{\rm p} \over \epsilon_{\rm r} \sigma_{\rm T} c}\, ,
\label{dMEdd}
\end{equation}
where $G$ is the gravitational constant, $\rho$ is the gas density, $c_s$ is the sound speed, $u$ is the gas velocity relative to the BH, $m_{\rm p}$ is the proton mass, $\epsilon_{\rm r}=0.1$ is the radiative efficiency of accretion, $\sigma_{\rm T}$ is the Thompson cross section, and $c$ is the speed of light. $\alpha=(\rho/\rho_0)^2$ for $\rho>\rho_0$ (1 otherwise) is an {\itshape ad hoc} boost parameter that models our inability to resolve the small scales of the interstellar medium where gas is denser and colder than the typical values reached with simulations of a kpc resolution \citep{booth&schaye09}.

We define the ratio $\chi=\dot M_{\rm BH}/ \dot M_{\rm Edd}$.
AGN feedback is modelled with two different forms: 
a {\itshape radio mode} with $\chi \le 0.01$: a kinetic bipolar outflow (jet) where mass, momentum, and energy are deposited within a small cylinder of radius $\Delta x$ and height $2 \Delta x$, and the jet velocity is $u_{\rm jet}=10^4$ km/s; 
 a {\itshape quasar mode} with $\chi > 0.01$: a thermal blast wave where energy is released in a pure thermal form within a bubble of radius $\Delta x$.
Each of the two modes of AGN feedback release a total amount of energy which is a fraction of the rest-mass accreted energy 
\begin{equation}
\dot E_{\rm AGN}=\epsilon_{\rm f} \epsilon_{\rm r} \dot M_{\rm BH}c^2\, ,
\label{E_BH}
\end{equation}
with $\epsilon_f=1$ in the radio mode and $\epsilon_f=0.15$ in the quasar mode.

Standard recipes for galaxy formation are also included with gas cooling, UV background heating, low efficiency star formation, supernova (SN) feedback and metal enrichment \citep[see][]{dubois&teyssier08winds}.

\section{Self-regulation of BH growth}

\begin{figure}
  \centering{\resizebox*{!}{5.8cm}{\includegraphics{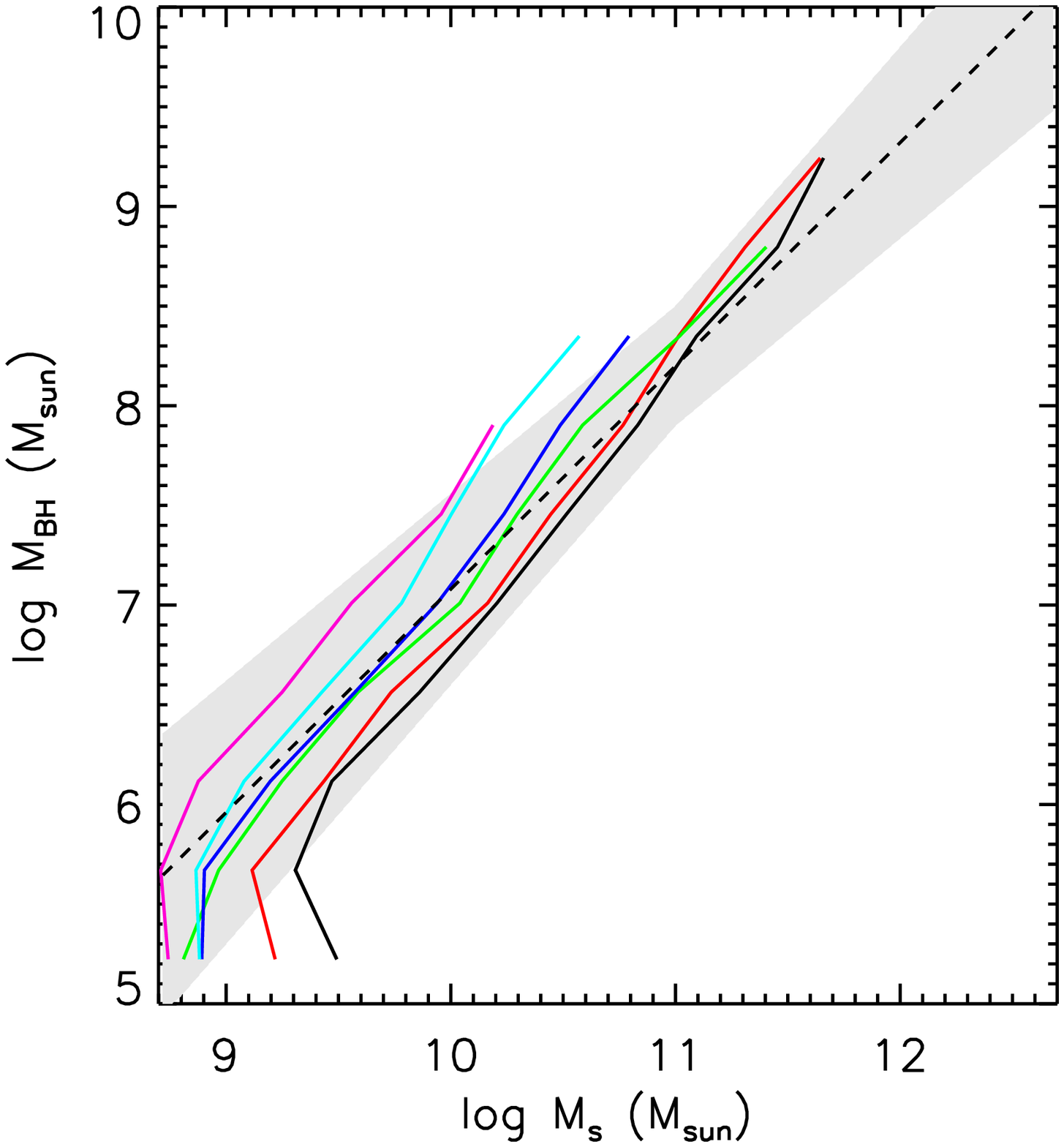}}\hspace{-1.5cm}}
  \centering{\resizebox*{!}{5.8cm}{\includegraphics{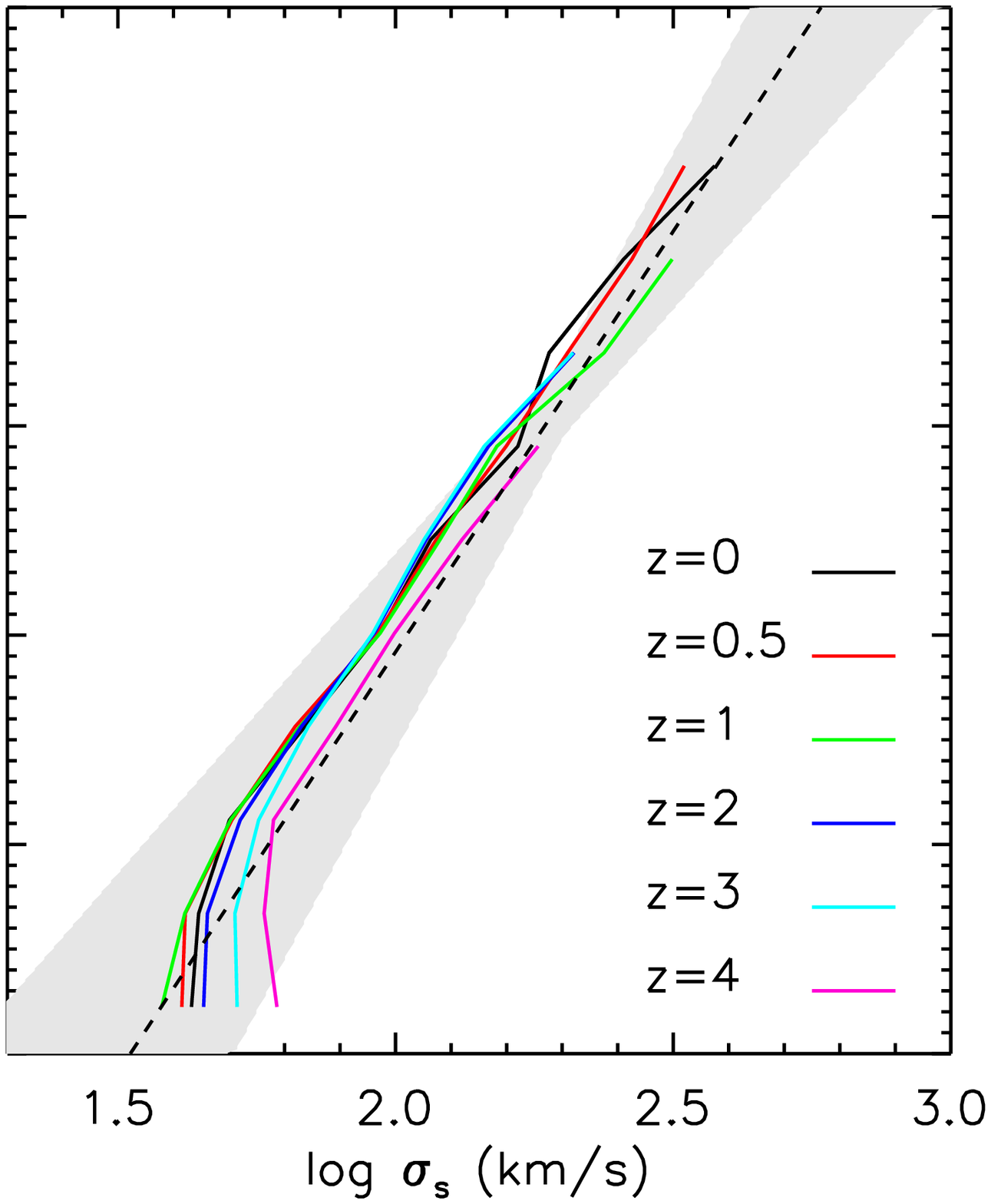}}}
  \caption{$M_{\rm BH}$/$M_{\rm s}$ (left panel) and $M_{\rm BH}$/$\sigma_{\rm s}$ (right panel) relations measured at different redshifts. Figure is taken from \cite{duboisetal11agnmodel}. Dashed lines with shaded areas correspond to observational fits at $z=0$ with $3\sigma$ errors.}
    \label{magorrian}
\end{figure}

Our AGN feedback model has several free parameters that we can tune independently from each other, such as the seed BH mass, the jet velocity , the  jet size, the efficiency of AGN energy release, etc. Varying these parameters one-by-one, we found one set of parameters that allow us to fit several observational constraints such as the BH mass versus stellar mass $M_{\rm BH}$/$M_{\rm s}$ relation, and the BH mass versus stellar velocity dispersion $M_{\rm BH}$/$\sigma_{\rm s}$ relation (see figure \ref{magorrian}), as well as the BH mass density and the cosmic SFR at $z=0$ \citep[see][for more details]{duboisetal11agnmodel}.

The $M_{\rm BH}$/$M_{\rm s}$ (or $M_{\rm BH}$/$\sigma_{\rm s}$) relation is a very good constraint of the AGN feedback efficiency $\epsilon_f$ required for self-regulation. Whatever the efficiency is, BHs grow sufficiently enough to unbind the cold surrounding gas while releasing the same amount of energy, and the $M_{\rm BH}$/$M_{\rm s}$ relation is shifted along $M_{\rm BH}$ by a factor $1/\epsilon_{\rm f}$. The $M_{\rm BH}$/$M_{\rm s}$ relationship shows some evolution with a decreasing ratio with redshift (figure 1). We find a slope from simulations $M_{\rm BH}$/$M_{\rm s}\propto (1+z)^\alpha$, $\alpha=0.42\pm 0.09$ compatible with observations $\alpha=0.68\pm0.12$ \citep{merlonietal10} at the $3 \sigma$ level.

We find a clear trend of the ratio of accretion $\chi$ decreasing with redshift. This behaviour is explained by the presence of large reservoirs of cold gas in galaxies at high redshift driven by cold filamentary inflows and numerous wet mergers. At lower redshift the accretion rate decreases because of the consumption of gas through the star formation process, the removal of gas by AGN outflows, and the shock-heating of atmospheres around massive galaxies. The consequence is that the quasar mode is the more common mode of AGN feedback in the young Universe, whereas the radio mode becomes more dominant when galaxies are evolving.

\section{Stopping the cooling catastrophe in galaxy clusters}

\begin{figure}
  \centering{\resizebox*{!}{5.8cm}{\includegraphics{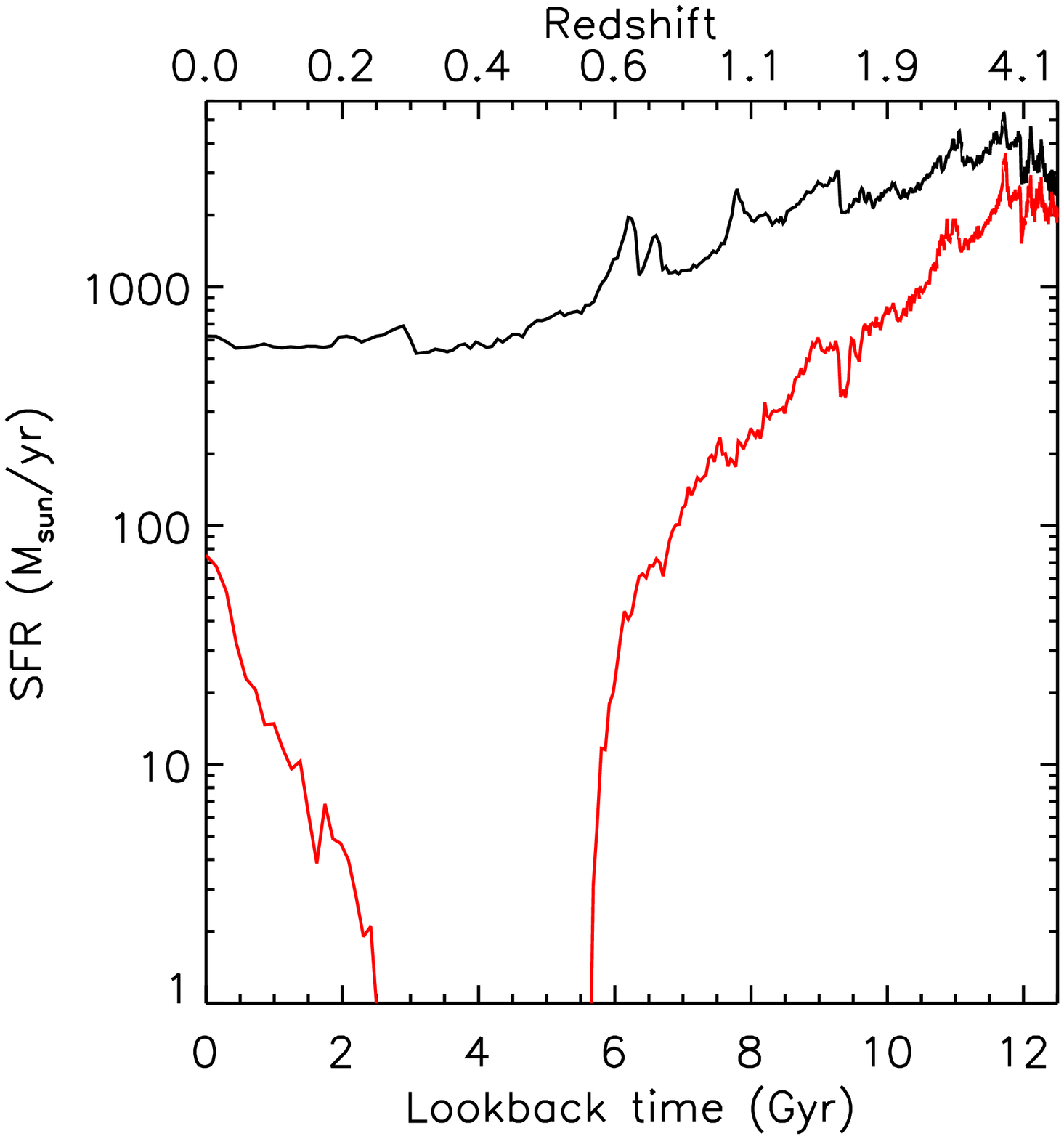}}}
  \centering{\resizebox*{!}{5.8cm}{\includegraphics{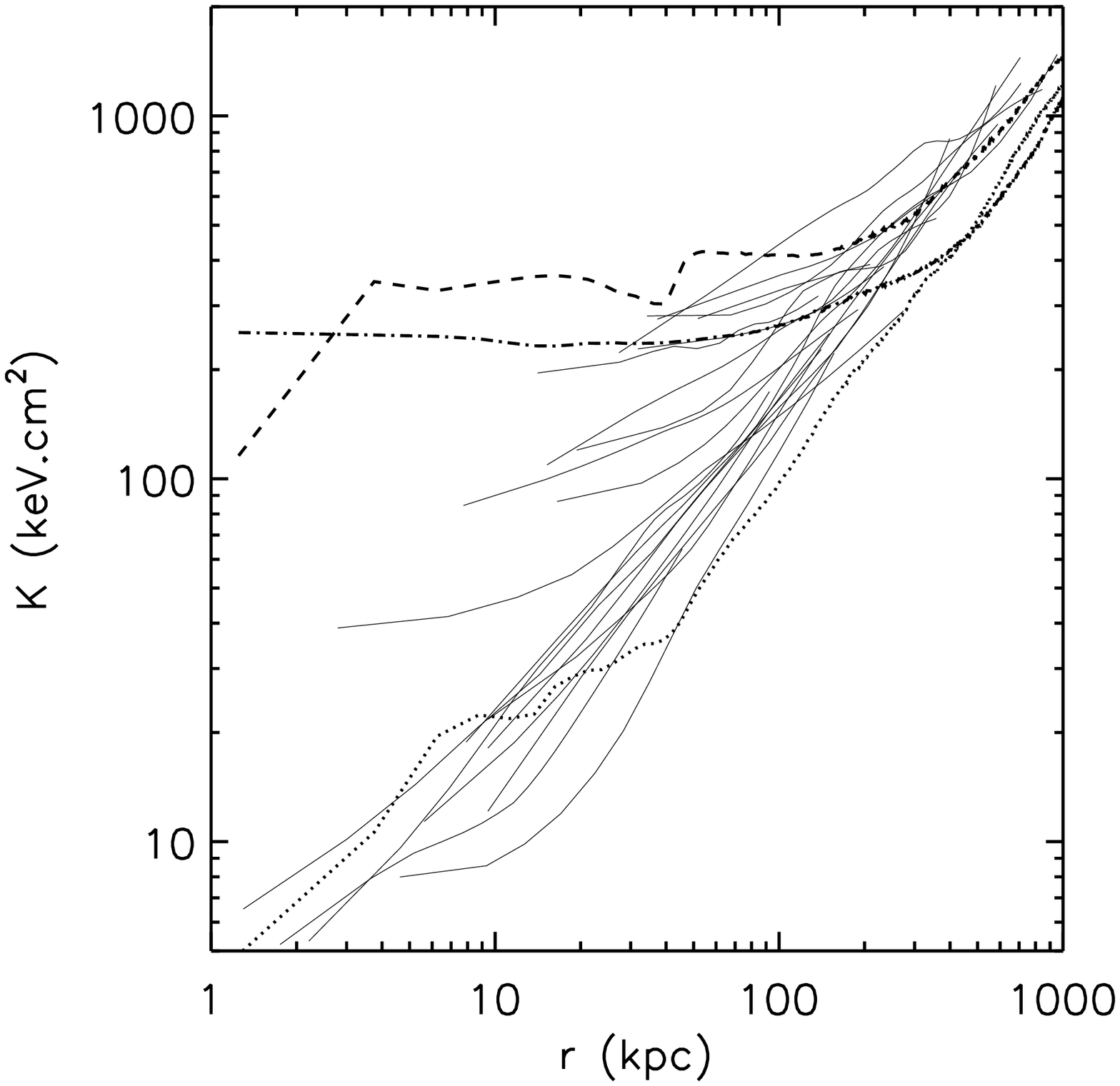}}}
  \caption{{\itshape (a)}: SFR vs time for the central elliptical galaxy of a simulated cluster of galaxies without AGN feedback (top line), and with AGN feedback (bottom line). {\itshape (b)}: Entropy profiles in a galaxy cluster at $z=0$ for a simulation with AGN feedback and metal cooling (dot-dashed line), or primordial H/He gas cooling with AGN feedback (dotted line) or without AGN feedback (dashed line). Solid lines are observations from \cite{sandersonetal09} for cool core and non-cool core clusters.}
    \label{cluster}
\end{figure}

In \cite{duboisetal10} we performed a simulation of a cosmological galaxy cluster with mass $M_{500}=2.10^{14}\,\rm M_{\odot}$ with only the radio mode of AGN feedback. We demonstrated that AGN feedback is able to stop the cooling catastrophe in the cluster by removing a lot of the cold gas in the bright central galaxy, and diminishes the amount of stars (figure~\ref{cluster}.(a)). We also showed that, by removing gas from the central parts of galaxy clusters, AGN feedback has an important impact on the concentration of material. The total mass profiles are less concentrated and the adiabatic contraction of DM is less pronounced \citep{teyssieretal11} 

AGN feedback prevents the low entropy material in the intra-cluster medium from cooling too fast (cooling catastrophe). Thus, entropy levels in the cores of galaxy clusters are lower and more realistic when AGN regulate cooling flows (figure~\ref{cluster}.(b)). One consequence is that we are able to obtain cool core and non-cool core entropy profiles in the same simulation. However the presence of metals released by SNe increases the cold gas accumulated at high redshift in galaxies, activates stronger bursts of AGN feedback with strong pre-heating, and increases the entropy levels in cluster cores in return \citep{duboisetal11}. 

\bibliography{author}

\begin{thebibliography}{}
\expandafter\ifx\csname natexlab\endcsname\relax\def\natexlab#1{#1}\fi
\expandafter\ifx\csname url\endcsname\relax
  \def\url#1{\texttt{#1}}\fi
\expandafter\ifx\csname urlprefix\endcsname\relax\def\urlprefix{URL }\fi
\providecommand{\eprint}[2][]{\url{#2}}

\bibitem[{{Booth} \& {Schaye}(2009)}]{booth&schaye09}
{Booth}, C.~M., \& {Schaye}, J. 2009, \mnras, 398, 53. \eprint{0904.2572}

\bibitem[{{Bower} et~al.(2006){Bower}, {Benson}, {Malbon}, {Helly}, {Frenk},
  {Baugh}, {Cole}, \& {Lacey}}]{boweretal06}
{Bower}, R.~G., {Benson}, A.~J., {Malbon}, R., {Helly}, J.~C., {Frenk}, C.~S.,
  {Baugh}, C.~M., {Cole}, S., \& {Lacey}, C.~G. 2006, \mnras, 370, 645.
  \eprint{arXiv:astro-ph/0511338}

\bibitem[{{Dubois} et~al.(2010){Dubois}, {Devriendt}, {Slyz}, \&
  {Teyssier}}]{duboisetal10}
{Dubois}, Y., {Devriendt}, J., {Slyz}, A., \& {Teyssier}, R. 2010, \mnras, 409,
  985. \eprint{1004.1851}

\bibitem[{{Dubois} et~al.(2011{\natexlab{a}}){Dubois}, {Devriendt}, {Slyz}, \&
  {Teyssier}}]{duboisetal11agnmodel}
--- 2011{\natexlab{a}}, ArXiv e-prints. \eprint{1108.0110}

\bibitem[{{Dubois} et~al.(2011{\natexlab{b}}){Dubois}, {Devriendt}, {Teyssier},
  \& {Slyz}}]{duboisetal11}
{Dubois}, Y., {Devriendt}, J., {Teyssier}, R., \& {Slyz}, A.
  2011{\natexlab{b}}, ArXiv e-prints. \eprint{1104.0171}

\bibitem[{{Dubois} \& {Teyssier}(2008)}]{dubois&teyssier08winds}
{Dubois}, Y., \& {Teyssier}, R. 2008, \aap, 477, 79. \eprint{arXiv:0707.3376}

\bibitem[{{Magorrian} et~al.(1998){Magorrian}, {Tremaine}, {Richstone},
  {Bender}, {Bower}, {Dressler}, {Faber}, {Gebhardt}, {Green}, {Grillmair},
  {Kormendy}, \& {Lauer}}]{magorrianetal98}
{Magorrian}, J., {Tremaine}, S., {Richstone}, D., {Bender}, R., {Bower}, G.,
  {Dressler}, A., {Faber}, S.~M., {Gebhardt}, K., {Green}, R., {Grillmair}, C.,
  {Kormendy}, J., \& {Lauer}, T. 1998, \aj, 115, 2285.
  \eprint{arXiv:astro-ph/9708072}

\bibitem[{{Merloni}(2010)}]{merlonietal10}
{Merloni}, ., A.~{et al} 2010, \apj, 708, 137. \eprint{0910.4970}

\bibitem[{{Sanderson} et~al.(2009){Sanderson}, {O'Sullivan}, \&
  {Ponman}}]{sandersonetal09}
{Sanderson}, A.~J.~R., {O'Sullivan}, E., \& {Ponman}, T.~J. 2009, \mnras, 395,
  764. \eprint{0902.1747}

\bibitem[{{Silk} \& {Rees}(1998)}]{silk&rees98}
{Silk}, J., \& {Rees}, M.~J. 1998, \aap, 331, L1.
  \eprint{arXiv:astro-ph/9801013}

\bibitem[{{Teyssier}(2002)}]{teyssier02}
{Teyssier}, R. 2002, \aap, 385, 337

\bibitem[{{Teyssier} et~al.(2011){Teyssier}, {Moore}, {Martizzi}, {Dubois}, \&
  {Mayer}}]{teyssieretal11}
{Teyssier}, R., {Moore}, B., {Martizzi}, D., {Dubois}, Y., \& {Mayer}, L. 2011,
  \mnras, 414, 195. \eprint{1003.4744}

\end{thebibliography}

\end{document}